\documentclass{article}
\usepackage[usenames]{color}
\usepackage{epsf}
\usepackage{amsmath}
\usepackage{epsfig,latexsym}
\usepackage{dirtytalk} 
\usepackage{amsmath,amsfonts,amstext,amssymb,amsbsy,amsopn,amsthm,eucal}
\usepackage{latexsym}
\usepackage{epsfig,latexsym}
\usepackage{indentfirst}
\usepackage{caption}

\title{Modelling Net Loan Loss with Bayesian and Frequentist Regression Analysis}
\author{Nathan Thomas Provost$\footnote{Student of Applied Mathematics and Statistics at Brown University. University Email: nathan\_provost@brown.edu}$}
\date{}

\begin{document}
\maketitle

\begin{center}
\textbf{Abstract}\\
\end{center}

\small

We create two distinct nonlinear regression models relating net loan loss (as an outcome) to several other financial and sociological quantities. We consider these data for the time interval between April $1^{st}$ 2011 and April $1^{st}$ 2020. We also include temporal quantities (month and year) in our model to improve accuracy. One model follows the frequentist paradigm for nonlinear regression, while the other follows the Bayesian paradigm. By using the two methods, we obtain a rounded understanding of the relationship between net loan losses and our given financial, sociological, and temporal variables, improving our ability to make financial predictions regarding the profitability of loan allocation.

\section*{Introduction}

Financial quantities are deeply interrelated with one another throughout all markets across the world. Loan loss in particular is a measure of interest that could benefit from more extensive analysis. There are a few financial quantities we should consider when investigating possible relationships. The federal funds rate (FFR) (which is the interest rate at which banks can loan each other their reserves) is an important variable, while the number of continued unemployment claims filed over a period of time is also a contributor. The average prime loan interest rate (APLIR) will also impact net loan loss. Finally, we  want to consider time (month and year) as a variable in our model, along with some sociological measures like gender and population size. An accurate regression model of net loan loss to total average loans for all banks in the U.S. is of great interest to both financial analysts and researchers, as it serves as an indicator of the country's financial well-being. Similarly, such a model would allow a bank to weigh its options when dealing with its clients, prompting the institution to allocate more or less loans based on the model's structure. Furthermore, we would gain a more comprehensive understanding of this temporal effect on loan loss. The passage of time is contextually important in finance, as it incorporates prior financial phenomena into the country's modern financial state. The extent to which this notion effects loan loss is not always clear, so we aim to investigate this overall effect. 
\newpage
\section*{Methods}

\normalsize

For our study, we use a mixture of population data and financial data, taken from  the federal reserve economics database (\cite{1} \cite{2} \cite{3}) and the U.S. census \cite{4}, respectively. We consider data from April 1$^{st}$ 2011 to April 1$^{st}$ 2020. Data points are reported quarterly (every 3 months), years are assigned discrete values (2011 = 1,..., 2020 = 10), and months are also assigned discrete values ([April 1$^{st}$ 2011] = 1, ..., [April 1$^{st}$ 2020] = 37). For the financial data (FFR, APLIR, and continued unemployment claims), the value reported for a given month is actually the average of all provided data from the previous month. This practice better incorporates contextual financial patterns, rather than simply using data for the first day of the month. The total population at the start of a given month is self-explanatory, while the sex ratio is simply the number of males at the start of a given month divided by the number of females. The data used herein covers the residential population in the United States along with all members of the armed forces, as detailed in the Census report \cite{4}. Any data entry for a given month or day that was left empty was omitted from our study, as it contributed no relevant information. All data was collected by its respective government agency (either the Federal Reserve or the U.S. census) in as exhaustive of a manner as possible, covering nearly all of the United States. Specific, rigorous details regarding data collection and data examination practices can be found within each of our sources (\cite{1} \cite{2} \cite{3} \cite{4}).\\

Additionally, as we will see, one of our models will require us to transform our data in order to better fit our data. These transformations are quite straightforward and only require us to scale our data down by a few orders of magnitude. We define each of the following variables to our data values: \say{Month} ($X_1$) represents the given month, \say{Year} ($X_2$)  represents the given year, \say{Loss} (L) represents the net loan losses to total average loans for all banks in the U.S., \say{TotalPop} ($X_3'$) represents the total population, \say{Ratio} ($X_4$) represents the male to female sex ratio, \say{APLIR} ($X_5$) represents the APLIR for the prior month, and \say{FFR} ($X_6$) represents the average federal funds rate for the prior month. From these definitions, we have:
\[\rm{AdjPop}=\frac{\rm{TotalPop}}{10^8} \ \ \ \rm{AdjClaims}=\frac{\rm{AVClaims}}{10^6} \ \ \ \rm{ExpClaims}=\textit{e}^{\rm{AdjClaims}}\]
The first two transformations arise out of a need to scale our data down, since the other values are significantly smaller than our \say{TotalPop} and \say{AVClaims} values. The necessity for \say{ExpClaims} will become apparent later on in our analysis, as our data may not initially follow a linear model very well. Similarly, Bayesian regression analysis is incredibly temperamental to large differences in the order of magnitude of different units, which is why we must scale down certain variables.\\

\newpage

\section*{Analysis}

We also conducted a few basic tests and provided a few essential statistics in our study. We present this basic collection of descriptive statistics (mean, standard deviation, median, minimum value, and maximum value) in Tables 1.1 and 1.2 and discuss them further on in our results. In addition, we will perform an analysis of variance on net loan loss grouped by month and net loan loss grouped by year. Our motivation in doing this is to make sure that including these variables in our model is necessary. Overall, our exploratory data analysis will help determine the necessity of certain elements in our models.\\

Our first model follows the traditional frequentist paradigm, but we encounter a problem when examining our AdjClaims values. There is an exponential relationship between Loss and AdjClaims ($X_7'$), which would mean that our linear model would be:
\[\mathbb{E}[L \mid X_1,...,X_6,X_7']=\hat{L}'=\beta_0 + \left[\sum_{i=1}^6 \beta_i X_i \right]+\beta_7e^{X_7'}\]
However, we have already defined $\rm{ExpClaims}=\textit{e}^{\rm{AdjClaims}}=\textit{e}^{X_7'}$, so if we denote ExpClaims by $X_7$, the model we will use is:
\[\mathbb{E}[L \mid X_1,...,X_7]=\hat{L}=\beta_0 + \sum_{i=1}^7 \beta_i X_i\]
This fulfills the structure of a typical linear model (though it is technically nonlinear with respect to AdjClaims). Using  this model, we will estimate the parameters ($\beta_0,...,\beta_7$) and interpret their values in the context of our model.\\

Our Bayesian model will have the same essential structure (with $X_7=e^{X_7'}$), but the underlying mathematical method through which we calculate the coefficients is different. In this method, we assume that parameters have their own distributions, which in turn means that we are interested in the posterior distributions of each of our regression coefficients. This also means our coefficients will have their own confidence intervals, which will be of interest to us. This estimation is done through the application of the relationship between Bayesian posteriors and priors and is repeated for each parameter. A more rigorous treatment of this method can be found in our sources (\cite{5} \cite{6} \cite{7} \cite{8}). \\

A few important constructions are essential to our Bayesian analysis, the first of which is the Region of Practical equivalence (ROPE) (\cite{7} \cite{8}). There are different ways to describe this region, but we shall adopt the traditional approach. Suppose $\sigma_L$ is the standard deviation of the variable \say{Loss}. Then, our ROPE would be:
\[\rm{ROPE}=\left[-\frac{\sigma_L}{10}, \ \frac{\sigma_L}{10}\right]\]
We will also consider the 89\% confidence interval (89\% is a traditional confidence level for Bayesian regression models). The midpoint of this interval serves as our parameter estimate. Finally, and most importantly, we consider the percentage of this interval that falls within the ROPE. This value will serve as a Bayesian analog to a frequentist p-value, but our rule for making decisions is slightly different. If 0\% of a given interval is found within the ROPE, then we conclude that there is a significant association between that parameter and the Loss variable. If 100\% (or very close to 100\%) of a given interval is found within the ROPE, then we conclude there is no significant association between the given parameter and Loss. If the percentage is between 0 and 100, then our conclusion is ambiguous.

\section*{Results}

\small

All of our data is analyzed using R (\cite{7} \cite{8} \cite{9} \cite{10}). We begin by examining our descriptive statistics (shown in Tables 1.1 and 1.2) and evaluating our preliminary analysis of variance tests. Important means and standard deviations are as follows (given in the form \say{mean (standard deviation)}): 15.3 (11.6) for exponentiated adjusted unemployment claims, 0.673 (0.781) for the effective federal funds rate, 3.79 (0.765) for the average prime loan interest rate, and 0.668 (0.387) for the net loan loss. Furthermore, our analysis of variance on the Loss variable grouped by month and year both yielded exceptionally small p-values ($2.47\times10^{-7}$ and $2.18\times10^{-7}$, respectively), which means that the mean Loss values grouped by month and year are \textbf{not} the same. This justifies our addition of these two variables to our models, which are detailed below.\\

Our data exhibits \textbf{no form} of multicollinearity and meets the requirements for linearity (after using ExpClaims), homoscedasticity, normality of errors, independence of errors, and null-mean of errors. For the Bayesian model, we also observed that the likelihood is roughly normal. Following the satisfaction of these assumptions, our frequentist model is as follows:
\[\hat{L}_F=-500+0.106X_1-0.0810X_2-18.5X_3+577X_4-0.995X_5+0.967X_6+0.0551X_7\]
This information follows directly from Table 2, and the overall model has an adjusted $R^2$ value of 0.971. Additionally, our Bayesian model is given by:
\[\hat{L}_B=-307.63+0.06X_1-0.08X_2-7.04X_3+340.98X_4-0.74X_5+0.72X_6+0.05X_7\]
This information corresponds to Table 3. The advantage of having both models present is that we can make more definitive conclusions about the associations between our independent variables and Loss. Subsequently, we will impose a strict rule for considering results significant, taking both models (p-values and percent in ROPE (PIROPE)) into account. \textbf{Independent variables with p-values less than 0.05 and PIROPE values equal (or very close) to 0 will be considered significant}. To further clarify, our confidence level for the frequentist model is 95\%, while our Bayesian model is set at 89\% (following the traditions of the appropriate literature). According to Table 2 and Table 3, the only variables meeting these criteria are APLIR ($X_5$), FFR ($X_6$), and ExpClaims ($X_7$), which means these three variables show a significant association with net loan loss for \textit{\textbf{both}} models. This leads us to the following interpretations (which are \textbf{each made} under the assumption that \textbf{all} other variables are held constant during a singular change). A one percent increase in the average prime loan interest rate for the prior month is significantly associated with a decrease of 0.995 and 0.74 percent in net loan loss to average total loans in the U.S. for a given month under the frequentist and Bayesian models, respectively. A one percent increase in the average effective federal funds rate for the prior month is significantly associated with an increase of  0.967 and 0.72 in net loan loss in the U.S. for a given month under the frequentist and Bayesian models, respectively. Finally, a one unit increase in exponentiated average continued unemployment claims in the U.S. for the previous month (adjusted to scale) is significantly associated with an increase of 0.0551 and 0.05 in net loan loss in the U.S. for a given month under the frequentist and Bayesian models, respectively. Finally, the p-values and PIROPE values for APLIR, FFR, and ExpClaims were $2.53\times10^{-2}$ and 0, $2.49\times10^{-2}$ and 0, and $2.49\times10^{-15}$ and 0, which indicate significance beyond a doubt under our model's guidelines and assumptions.
\section*{Discussion and Conclusion}
\small
Evidently, continued unemployment claims are strongly associated with an increase in net loan loss to total average loans for all banks in the U.S. The p-value for this variable is much smaller than the other two p-values, which signifies a more significant relationship than the other two variables. This means that the number of unemployment claims can be used as a means of predicting large increases in loan loss. It is important to note, however, that this association is exponentially defined, and not linearly defined in the traditional sense. Overall, our model takes several key financial and sociological quantities into account with considerable accuracy. Clearly, the linear model is extremely well-fitting (as shown by our $R^2$ value of 0.971), and our Bayesian model closely follows this trend. We could certainly include further variables to strengthen our model, perhaps trying to include more socio-political factors, but all in all, the models seem to collectively show the three most important variables. A fair criticism of this study would be that the method by which we label variables significant is far too strict, as they need to demonstrate significance for the Bayesian model at 89\% confidence \textbf{\textit{and}} the frequentist model at 95\% confidence. We could certainly investigate the other variables present, but the fact that average prime loan interest rate, effective federal funds rate, and exponentiated adjusted unemployment claims pass significance thresholds for both models indicates that their influence is exceptionally notable. Future studies may wish to include more demographic variables, in order to compare different parts of the U.S. to one another, or alternatively, incorporate financial laws between states into the model. Nonetheless, we have detected a crucial relationship between net loan loss and unemployment claims, which has strengthened and expanded our financial understanding and predictive abilities.\\

However, our data is straddled with many of the limitations that come with using government data, which should be noted. Census data suffers (as it always has) from under-representation of certain groups and over-representation of others (this is mentioned in the methodology linked in the legend from our source \cite{4}). As such, the data is not entirely accurate as we have assumed, just as not every inhabitant of the U.S. is ever truly counted. Similarly, our Bayesian model is not easily generalizable, since large differences in units can drastically reduce the model's accuracy. This is why we had to adjust the population and unemployment claims data. Future studies should aim to avoid such scaling or carefully re-scale the data, in order to preserve accuracy. Finally, though the sociological variables we considered (sex ratio and total population size) were not significantly associated with net loan loss, there are other sociological factors that were not considered that should be. Levels of diversity in the United States should be added to the model to enhance its holistic accuracy and provide a more rounded picture. Despite these limitations, the relationship shown by our model is still of immense value to analysts and researchers due to its predictive power and general accuracy, and future research should aim to bolster these qualities.

\section*{Acknowledgements}

I am indebted to Dr. Shira Dunsiger (Associate Professor of Behavioral and Social Sciences at the Brown University School of Public Health) for her advice and constructive criticism regarding this paper.

\newpage
\normalsize
\section*{Tables}

\begin{figure}[!ht]
    \centering
    \includegraphics[width=7.5cm]{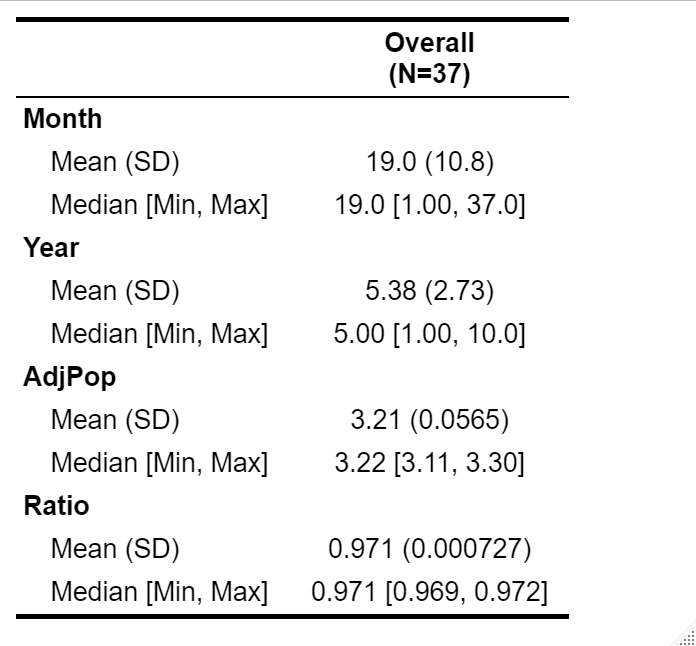}
    {\caption*{Table 1.1: Descriptive Statistics for Population Data and Temporal Data}}
\end{figure}

\begin{figure}[!ht]
    \centering
    \includegraphics[width=7.5cm]{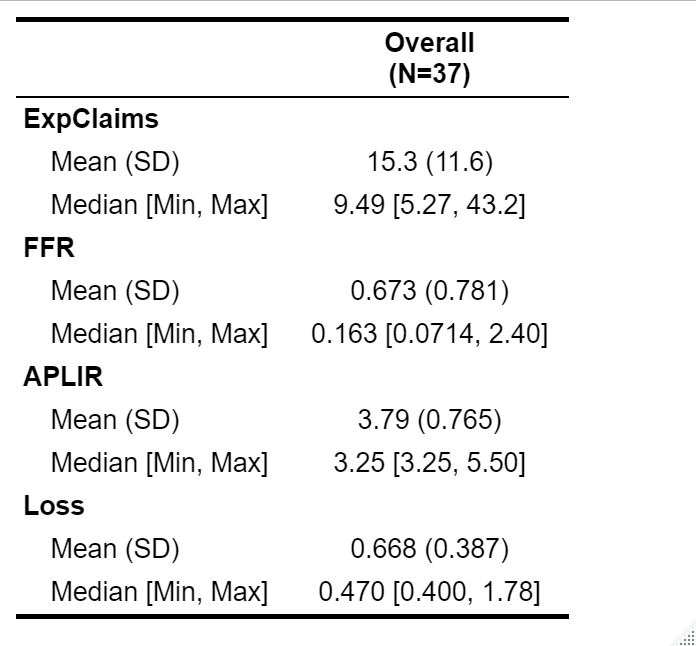}
    {\caption*{Table 1.2: Descriptive Statistics for Financial Data}}
\end{figure}
\newpage
\noindent\textbf{Table 2}\\
\begin{verbatim}
 term       | estimate  | std.error | statistic | p.value
----------------------------------------------------------
 (Intercept)|   -500.   |  275.     |   -1.82   | 7.96e- 2
 Month      |    0.106  |  0.0519   |    2.05   | 4.98e- 2
 Year       |   -0.0810 |  0.0396   |   -2.04   | 5.01e- 2
 ExpClaims  |    0.0551 |  0.00363  |    15.2   | 2.49e-15
 APLIR      |   -0.995  |  0.422    |   -2.36   | 2.53e- 2
 FFR        |    0.967  |  0.409    |    2.37   | 2.49e- 2
 AdjPop     |   -18.5   |  13.4     |   -1.38   | 1.78e- 1
 Ratio      |    577.   |  323.     |    1.79   | 8.45e- 2
\end{verbatim}
\noindent\textbf{Table 3}\\
\begin{verbatim}
Parameter   |  Median |       89% CI      |      ROPE      | % in ROPE |
------------------------------------------------------------------------
(Intercept) | -307.63 | [-667.56,  52.00] | [-0.04, 0.04]  |        0% |
Month       |    0.06 | [  -0.00,   0.12] | [-0.04, 0.04]  |    27.55% |
Year        |   -0.08 | [  -0.14,  -0.01] | [-0.04, 0.04]  |    13.51% |
ExpClaims   |    0.05 | [   0.05,   0.06] | [-0.04, 0.04]  |        0% |
APLIR       |   -0.74 | [  -1.39,  -0.14] | [-0.04, 0.04]  |        0% |
FFR         |    0.72 | [   0.12,   1.34] | [-0.04, 0.04]  |        0% |
AdjPop      |   -7.04 | [ -22.56,   8.94] | [-0.04, 0.04]  |     0.34% |
Ratio       |  340.98 | [ -79.36, 749.00] | [-0.04, 0.04]  |        0% |

\end{verbatim}

\begin{thebibliography}{9}

\bibitem{1} Federal Reserve Bank of New York. (2021, April 5). Effective Federal Funds Rate. FRED. https://fred.stlouisfed.org/series/EFFR. 
\bibitem{2} Board of Governors of the Federal Reserve System (US). (2021, April 5). Bank Prime Loan Rate. FRED. https://fred.stlouisfed.org/series/DPRIME. 

\bibitem{3} U.S. Employment and Training Administration. (2021, April 1). Continued Claims (Insured Unemployment). FRED. https://fred.stlouisfed.org/series/CCSA. 

\bibitem{4} US Census Bureau. (2020, June 17). National Population by Characteristics: 2010-2019. The United States Census Bureau. https://www.census.gov/data/tables/time-series/demo/popest/2010s-national-detail.html\#par\_textimage\_98372960. 

\bibitem{5} Halimi, A., Mailhes, C., \& Tourneret, J.Y. (2017). Nonlinear Regression Using Smooth Bayesian Estimation. Toulouse, France. 

\bibitem{6} Katz, D., Azen, S., \& Schumitzky, A. (1981). Bayesian Approach to the Analysis of Nonlinear Models: Implementation and Evaluation. Biometrics, 37(1), 137-142. doi:10.2307/2530529

\bibitem{7} Makowski, D., Ben\-Shachar, M. S., \&; Lüdecke, D. (2019). Region of Practical Equivalence (ROPE). • bayestestR. https://easystats.github.io/bayestestR/articles/region\_of\_practical\_equivalence.html.

\bibitem{8} Makowski, D., Ben-Shachar, M. S., \&; Lüdecke, D. (2019). 1. Initiation to Bayesian models. • bayestestR. https://easystats.github.io/bayestestR/articles/example1.html\#all-with-one-function-1. 

\bibitem{9} R Core Team (2019). R: A language and environment for statistical computing. R Foundation for Statistical Computing, Vienna, Austria. URL https://www.R-project.org/.

\bibitem{10} David Robinson, Alex Hayes and Simon Couch (2020). broom: Convert Statistical Objects into Tidy Tibbles. R package version 0.7.1. https://CRAN.R-project.org/package=broom

\end{thebibliography}
\end{document}